\documentclass[aps,amsmath,amssymb, preprint]{revtex4}

\usepackage{amsfonts}
\usepackage{amsmath}
\usepackage{amssymb}
\usepackage{graphicx}
\usepackage{color}

\newcommand{\kvec}{\mathbf{k}}

\newcommand{\del}{\nabla}
\newcommand{\beq}{\begin{equation}}
\newcommand{\eeq}{\end{equation}}
\newcommand{\bsp}{\begin{split}}
\newcommand{\esp}{\end{split}}
\newcommand{\bpm}{\begin{pmatrix}}
\newcommand{\epm}{\end{pmatrix}}

\newcommand{\eq}[1]{Eq. [\ref{#1}]}
\newcommand{\fig}[1]{Fig.[\ref{#1}]}

\newcommand{\dae}[2]{\frac{\partial #1}{\partial #2}}

\begin{document}

\title{Signatures of the Berry curvature in the frequency dependent interlayer magnetoresistance in tilted magnetic fields}

\author{Anthony R. Wright}
\email{a.wright7@uq.edu.au}
\affiliation{School of Mathematics and Physics, University of Queensland, Brisbane, 4072 Queensland, Australia}
\author{Ross H. McKenzie}
\affiliation{School of Mathematics and Physics, University of Queensland, Brisbane, 4072 Queensland, Australia}

\date{\today}

\begin{abstract}
We show that in a layered metal, the angle dependent, finite frequency, interlayer magnetoresistance is altered due to the presence of a non-zero Berry curvature at the Fermi surface. At zero frequency, we find a conservation law which demands that the `magic angle' condition for interlayer magnetoresistance extrema as a function of magnetic field tilt angle is essentially both field and Berry curvature independent. In the finite frequency case, however, we find that surprisingly large signatures of a finite Berry curvature occur in the periodic orbit resonances. We outline a method whereby the presence and magnitude of the Berry curvature at the Fermi surface can be extracted.  
\end{abstract}


\maketitle

\section{Introduction}
Berry's geometric phase and the corresponding Berry curvature, alter many of the electronic and transport properties of materials \cite{niurev}. A notable example is the set of Hall effects: the TKNN invariant relevant to quantum Hall insulators \cite{TKNN}, the quantum anomalous Hall effect \cite{haldane,AHE}, the $Z_2$ indices for the quantum spin Hall effect \cite{Z2}, and the related anomalous Nernst effect \cite{nernst}. These can all be connected to the Berry curvature and the corresponding first Chern number \cite{chern}. However, experimental detection of many of these effects, and measurement of the Berry curvature itself, is difficult.


In a tilted magnetic field, when measuring interlayer resistivity, angle dependent magnetoresistance oscillations (AMRO) are observed in layered metals as a function of the tilt angle. The experimental setup is shown in \fig{setup}. The oscillations can be viewed as an Aharanov-Bohm interference pattern due to the magnetic flux enclosed by the possible cyclotron orbits between layers, which are shifted relative to each other by the in-plane component of the magnetic field \cite{cooperAB}. AMRO have now been used extensively to extract information about the Fermi surface properties of layered metals \cite{hussey, mackenzie,organics}. AMRO has become crucial to the study of overdoped cuprates \cite{hussey} and organic superconductors \cite{organics}, and the study of their phase diagrams on varying temperature, pressure, and doping.  While the analysis of magnetic quantum oscillations in topological insulators is a subtle issue \cite{qo1,qo2,qo3,ando,tonyross}, it seems that these experiments are a powerful tool in probing the topological properties of quantum  matter \cite{tony2, moorerev}. AMRO experiments on topological insulators have recently begun \cite{andoExpt, andoExpt2, andoExpt3}. By performing AMRO experiments at finite frequency, periodic orbit resonances \cite{POR0, POR} can be observed and used to extract even more information about the Fermi surface properties of a layered metal, for instance mapping out its Fermi velocity \cite{Kovalev, Hill2}, and investigating the doping and temperature dependence of Fermi surfaces \cite{organics07}. Periodic orbit resonances up to seventh order have been used in quasi-one dimensional and quasi-two dimensional organics to comprehensively map out anisotropic Fermi surfaces\cite{oshima}.

One striking aspect of AMRO experiments in layered systems is their seemingly universal behavior, specifically in the appearance of `magic angles'  -- the magnetic field tilt angles at which the interlayer resistance is an extremum, which are independent of the magnitude of the magnetic field and temperature. In this Article, we study the effect of the Berry curvature on AMRO. We demonstrate a simple conservation law which demands that the magic angle condition, to lowest order, is unaffected by the Berry curvature at the Fermi surface. To higher order, we find that the Berry curvature on the Fermi surface gives a distinct signature in the magnetic field strength dependence of the resistance extrema. This effect is \emph{very} small. To aid in the experimental measurement of the Berry curvature at the Fermi surface, we study AMRO in a time dependent electric field, and calculate the frequency dependence of the AMRO. In this case we find that the periodic orbit resonances \cite{MMfreq} are substantially altered by the Berry curvature, and thus form a viable probe of the Berry curvature at the Fermi surface.

\section{Angle dependent magnetoresistance oscillations and magic angles}
\begin{figure}[tbp]
\centering\includegraphics[width=8cm]{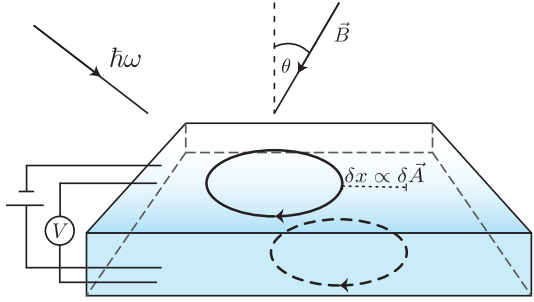}
\caption{The experimental setup: measuring the interlayer magnetoresistance in a tilted magnetic field and a time-dependent electric field. The out-of-plane component of the magnetic field ($B_\perp = |\vec{B}|\cos\theta$) initiates cyclotron orbits along the intra-plane Fermi surfaces within the layers of the system. The in-plane component of the magnetic field ($B_\parallel = |\vec{B}|\sin\theta$) shifts the Fermi surfaces of subsequent layers with respect to each other, which is equivalent to a real space shift shown schematically here. This shift is proportional to the vector potential $\vec{A}$. The time dependent electric field can be from an external light source ($\hbar \omega$ in the figure), or a time dependence in the applied voltage, $V$.}
\label{setup}
\end{figure}

It has been shown that for both quasi-two dimensional and quasi-one dimensional systems, whether there are just two layers or a superlattice \cite{moses}, irrespective of the form of the intralayer Hamiltonian and whether the interlayer transport is coherent or weakly incoherent \cite{smith,kennet}, that the interlayer conductivity \emph{always} takes the same form. At tilt angle $\theta$, as defined in \fig{setup}, and at electric field frequency $\omega$, this is given by \cite{moses, MMfreq}

\beq
\frac{\sigma_\perp(\theta,\omega)}{\sigma_{\perp}(0)} = \sum_{\nu=-\infty}^\infty \frac{J_\nu(\mu)^2}{1 + (\omega\tau - \nu\omega_c  \tau \cos\theta)^2},
\label{sig}
\eeq  
for a circular intralayer Fermi surface, where $\mu = ck_F \tan \theta$, in which $c$ is the layer separation, and $k_F$ the Fermi wavevector. $J_\nu$ is the modified Bessel function of the $\nu^{\mathrm{th}}$ kind, $\omega_c^\alpha$ is the cyclotron frequency, and $\tau$ is the relaxation time. 

For large fields ($\omega_c^\alpha \tau\gg1$), the first term in \eq{sig} dominates such that the magic angle condition, at which the interlayer magnetoresistance is a local maximum, satisfies 
\beq
ck_F \tan \theta_n = \pi(n-1/4),
\label{MAold}
\eeq
for $n\in \mathcal{Z}$, when $c k_F \tan\theta_n\gg1$, such that we can take the asymptotic form of the Bessel function. 

In this Article, we show that the Berry curvature does not modify these expressions. Rather, we find a strict conservation law that prevents such a modification to lowest order in the series, \eq{sig}. This is rather surprising given that the Berry curvature does modify the electron orbits on the Fermi surface. On the other hand, we introduce a different measure of conductivity extrema, corresponding to periodic orbit resonances in a time dependent electric field. These are found to appreciably shift as a function of tilt angle and magnetic field magnitude due to the presence of a Berry curvature, and thus form a tangible probe of the Berry curvature on the Fermi surface.

\section{Calculating AMRO for layered systems.} 
Consider a general superlattice Hamiltonian, which is an infinite series of identical stacked two-dimensional layers. The layers are weakly coupled by an overlap integral $t_\perp$ and contain two-species which can be any real or pseudo-spin such as a sublattice or on-site orbitals. The full Hamiltonian is given by $\hat{H} = \sum_{\vec{k}} H(\vec{k})c_{\vec{k}}^\dag c_{\vec{k}}$, where $c_{\vec{k}}$ is a spinor in the relevant pseudo-spin basis, $\vec{k}$ is a Block wavevector, and

\beq
H(\vec{k}) = d_0(\vec{k}_\parallel) \sigma_0 + \vec{d}(\vec{k}_\parallel)\cdot\vec{\sigma} - 2t_\perp \cos(k_z)\sigma_0,
\label{HSL}
\eeq
in which $\sigma_0$ is the $2\times2$ identity matrix, and $\sigma_{x,y,z}$ are the Pauli matrices. For small $t_\perp$, the Fermi surface is approximately a cylinder. The eigenvalues are

\beq
\epsilon_\alpha(\vec{k}) = d_0(\vec{k}_\parallel) +\alpha |\vec{d}(\vec{k}_\parallel)| - 2t_\perp\cos(k_z),
\label{eps}
\eeq
where $\alpha = \pm1$. The intralayer part of this Hamiltonian describes graphene \cite{graphene} and its cousins \cite{silicene}, topological insulator surface states \cite{KM,BHZ}, spintronics systems \cite{spintronics}, and some unconventional triplet superconductors (e.g., the $p+ip$ variety) \cite{zhangrev}. In graphene, for example, $\vec{d}(\vec{k}) = (\pm k_x,k_y,0)$ near the $K_\pm$ points.

We now solve the relevant Boltzmann equation within the relaxation time approximation, as has been done previously \cite{moses}. However, we include the Berry curvature  \cite{berry}

\beq
\Omega_\alpha(\vec{k}) = i\langle \nabla_{\vec{k}} u_{\vec{k},\alpha}|\times |\nabla_{\vec{k}} u_{\vec{k},\alpha}\rangle,
\eeq
where $|u_{\vec{k},\alpha}\rangle$ are the eigenvectors of \eq{HSL}.

The semi-classical equations of motion, including the Berry curvature, are given by ($\hbar = 1$) \cite{motion}

\beq
\bsp
&\frac{d\vec{r}}{dt} = \dae{\epsilon_\alpha}{\vec{k}} - \frac{d\vec{k}}{dt}\times\vec{\Omega}_\alpha(\vec{k})\\
&\frac{d\vec{k}}{dt} = -e\vec{E} - e\frac{d\vec{r}}{dt}\times\vec{B}.
\end{split}
\label{niu}
\eeq
Combining the two, we can remove the real space coordinate, finding

\beq
(1 + e\vec{B}\cdot\vec{\Omega}_\alpha(\vec{k}))\frac{d\vec{k}}{dt}= -e\vec{E} - e\dae{\epsilon}{\vec{k}}\times\vec{B} + e\vec{\Omega}_\alpha(\vec{k})(\vec{B}\cdot \frac{d\vec{k}}{dt}).
\label{kdot}
\eeq
Solving for the intralayer equations of motion when $\vec{E} = \vec{0}$, assuming the interlayer hopping $t_\perp$ is small, we obtain $k_y + ik_x = k_Fe^{-i\omega_c^\alpha  t}$, where the semi-classical cyclotron frequency is given by

\beq
\omega^\alpha_{c} = \frac{eB_\perp}{k_F(1 + eB_\perp\Omega_\alpha(k_F))}\dae{\epsilon_\alpha(k_F)}{k_F},
\label{wc}
\eeq
which is clearly a function of the Berry curvature $\Omega_\alpha(k_F)$. 

Solving for the interlayer wavevector $k_z(t)$, we first observe from the second expression in \eq{niu}, that in the absence of an electric field, $d\vec{k}/dt\cdot\vec{B} = 0$, and thus $\vec{k}\cdot\vec{B}$ is a constant of motion. Expanding this component-wise, and setting $B_y=0$, we can write

\beq
k_z(t) = k_z(0) + \frac{B_x}{B_z}\bigl(k_x(t) - k_x(0)\bigr).
\label{kzt}
\eeq
Writing $\vec{B} = |\vec{B}|(\sin\theta, 0, \cos\theta)$, and substituting the expression for $k_x(t)$, we obtain

\beq
k_z(t) = k_z(0) + k_F\tan\theta\sin(\omega_c^\alpha t).
\eeq



From this, we can calculate the interlayer conductivity, in an analogous way as was performed previously in the absence of a Berry curvature (see Ref. [\onlinecite{moses}], Eq. 25ff),  the difference in the two results being that here the cyclotron motion depends upon the Berry curvature, as shown in \eq{wc}. Therefore, we can immediately present our result for the interlayer conductivity, \eq{sig}.

We see then, that from the conservation law leading to \eq{kzt}, the magic angle condition, which arises from the first term in the series for the interlayer conductivity, \eq{sig}, \emph{cannot} be altered by the Berry curvature.

Remarkably, an identical result is obtained for bilayer systems \cite{moses, cooper}. The calculation in this case does not differ from that presented elsewhere. For the sake of completeness, we briefly sketch the calculation here.

Consider a general bilayer Hamiltonian with layer index denoted $i = 1,2$, given by

\beq
\hat{H} = \sum_{\vec{k}}\biggl[ \sum_{i=1}^2 H_i(\vec{k}) c_{i,\vec{k}}^\dag c_{i,\vec{k}}  + t_\perp c_{1,\vec{k}}^\dag c_{2,\vec{k}} + h.c\biggr]
\eeq
where $\vec{k}$ is now in-plane only and is conserved upon interlayer hopping. We use two identical intralayer Hamiltonians of the form $H_i(\vec{k}) = d_{0}(\vec{k}) + \vec{d}(\vec{k})\cdot \vec{\sigma}$. The intralayer Hamiltonians each induce a Berry curvature $\Omega(\kvec)$. 

We follow Cooper and Yakovenko's approach to the problem \cite{cooper}, namely, incorporating the Peierls' substitution into the interlayer hopping integral, $t_\perp(t) \rightarrow t_\perp e^{i\frac{e}{\hbar} \int_0^c B_\parallel y(t) dz}$, where the two layers are located at $z = [0, c]$. Using the gauge $\vec{A}(t) = B_\parallel y(t) \hat{z}$, associated with the component of the field parallel to the layers, we obtain 

\beq
t_\perp \rightarrow t_\perp e^{i c k_F \tan(\theta)\sin(\omega_c^\alpha  t)}
\eeq
The Kubo formula for interlayer conductivity can be calculated by phase averaging the square of $t_\perp$ \cite{cooper}. Including the relaxation time, this leads to the interlayer conductivity \eq{sig}. 


\section{Berry curvature fingerprints in higher order corrections to the magic angle condition}
Experimental measures of the Berry curvature at the Fermi surface are elusive. Spurred on by this difficulty, we now explore the second order term in the conductivity \eq{sig}. In order to experimentally observe inter- or intra-layer magnetic quantum oscillations, the relaxation rate must be at least of the same order as the cyclotron frequency, and would ideally be much smaller. In smaller fields, such that $\omega_c^\alpha\tau \sim 1$, the second order term in \eq{sig} becomes relevant. Expanding this term, we obtain for small $eB_\perp\Omega$,

\beq
\frac{\sigma_\perp(\theta)}{\sigma_{\perp}(0)} \approx J_0(\mu)^2 + 2 \frac{J_1(\mu)^2(1+2e|B|\Omega \cos\theta)}{1 + (\omega^0_c  \tau \cos\theta)^2},
\label{sig2}
\eeq  
where $\omega_c^0$ is the cyclotron frequency, \eq{wc}, with zero Berry curvature, $\Omega = 0$, and as we are interested here in the zero frequency result, we have set $\omega=0$. Taking the asymptotic form of the Bessel functions, we obtain

\beq
\frac{\sigma_\perp(\theta)}{\sigma_{\perp}(0)} \approx \frac{2}{\pi\mu}\biggl[\cos(\mu(\theta) - \pi/4)^2 - 2 \frac{\cos(\mu(\theta)+\pi/4)^2(1+2e|B|\Omega \cos\theta)}{1 + (\omega^0_c  \tau \cos\theta)^2}\biggr],
\label{sig22}
\eeq  
where we have included the angle dependence of $\mu(\theta) = ck_F\tan\theta$ for emphasis. A simple analytic expression for the extrema of the conductivity in this case does not exist. However we can see that the conductivity is now a function of magnetic field strength multiplied by Berry curvature.

Unfortunately, as we shall see later, this higher order correction due to the Berry curvature in most realistic systems is miniscule. Furthermore, since the term $\omega_c^0 \propto |B|\cos\theta$, the second order correction above varies with the field magnitude in a rather complicated way, making a meaningful extraction of the Berry curvature at the Fermi surface quite subtle. 

\section{Mapping out the Berry curvature with periodic orbit resonances}
\begin{figure}[tbp]
\centering\includegraphics[width=8cm]{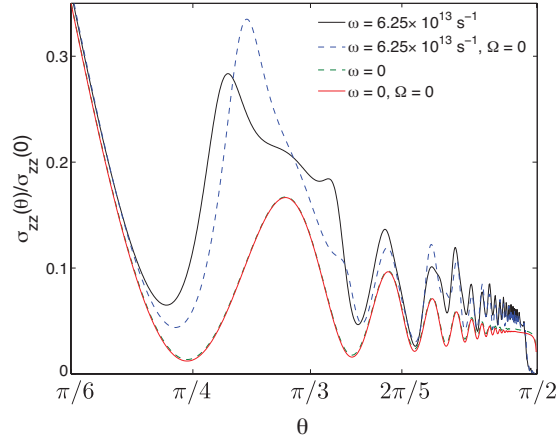}
\caption{A typical angle dependent interlayer conductivity curve at both zero and finite electric field frequency, and with both zero and non-zero Berry curvature. For the zero frequency case, the Berry curvature does not alter the location of the conductivity extrema, and only slightly alters the magnitude of the conductivity. At finite frequency however, the Berry curvature alters the locations of the extrema, as well as their amplitude significantly, thus leaving a substantial fingerprint that can be used to measure the Berry curvature at the Fermi surface. System parameters are $v_F = 10^5$ ms$^{-1}$, $\tau = 2\times 10^{-13}$ s, $c = 5\times10^{-8}$ m, $k_F = 5\times10^7$ m$^{-1}$, $\omega = 6.25\times10^{13}$ s$^{-1}$, $g_s = 5$, $B = 20$ T. The Berry curvature to magnetic length ratio with these parameters is $\Omega/l_B^2 \approx 0.06$.}
\label{freqfigfull}
\end{figure}

Including a time dependent electric field, we find that the structure of the interlayer conductivity develops additional structure. In \fig{freqfigfull} we have shown the interlayer conductivity, as a function of tilt angle, at both zero and finite frequencies, and both with and without Berry curvature. Most striking, together with the additional structure, is that the Berry curvature, which is almost completely unnoticeable in the zero frequency case,  alters the interlayer conductivity. In particular, the location of the extrema have moved significantly due to the Berry curvature.  

\begin{figure}[tbp]
\centering\includegraphics[width=8cm]{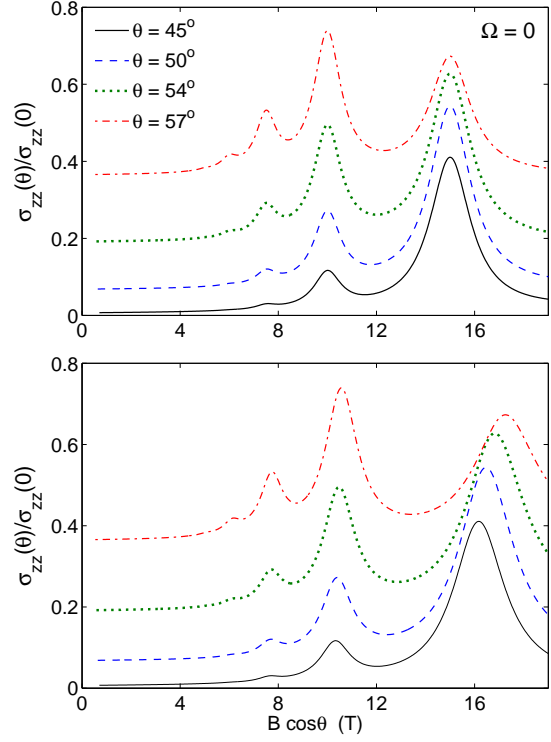}
\caption{The periodic orbit resonances at different magnetic field tilt angles $\theta$, plotted as a function of the z-component of the magnetic field (i.e., $B_z = |\vec{B}|\cos\theta$), at fixed electric field frequency $\omega$. Without a Berry curvature (upper pane), the resonances always occur at the same value of $B_z$. A finite Berry curvature (lower pane) alters this condition, through the change in cyclotron frequency, \eq{wc}, such that the Berry curvature on the Fermi surface can be measured. For example,  over the $12^o$ change in tilt angle shown, the resonance near 16-17 T shifts by $\approx1.1$ T, and the resonance near 10 T shifts by $\approx 0.2$ T. Material parameters are the same as those used in \fig{freqfigfull} The traces have been offset from each other for clarity.}. 
\label{resfig}
\end{figure}

Inspecting the denominators of the series expansion for the conductivity in \eq{sig}, we see that resonances occur when 

\beq
\omega = \nu\omega_c^\alpha\cos\theta,\,\,\,\,\nu\in\mathcal{Z}.
\label{rescond}
\eeq
 These resonances, which occur on top of the oscillating signal from the Bessel functions in the conductivity, lead to the complicated pattern observed in \fig{freqfigfull}. In order to extract information about the Berry curvature from this signal, we expand the resonance condition, \eq{rescond}, in terms of the magnetic field strength, and the tilt angle $\theta$, and find that the $\nu^{\mathrm{th}}$ resonance occurs when

\beq
\frac{\cos\theta}{l_B^2}\biggr|_\nu = \frac{\nu\omega k_F}{\partial_{k_F}\epsilon_\alpha(k_F)}\biggl(1 + \frac{\Omega(k_F)}{l_B^2}\cos\theta\biggr),
\label{rescond2}
\eeq 
where $l_B^2 = \hbar/eB$ is the magnetic length. Notably, if the Berry curvature, $\Omega(k_F)$, is zero, then the periodic resonances in the conductivity will always occur at the same values of $\cos\theta/l_B^2$, irrespective of the tilt angle or magnetic field strength. In \fig{resfig} we have shown the resonances for two systems, one with a Berry curvature, and one without. As expected from the above expression, as the tilt angle is varied, the resonances remain constant \emph{unless there is a finite Berry curvature}. We note that the effect is significantly greater than in the zero frequency case, and so is much more readily observed.  

\section{Sample materials and experimental questions}
The expressions obtained in the previous sections are for a general $SU(2)$ intralayer Hamiltonian with weak interlayer tunnelling amplitude. The prototypical Hamiltonians of such a form that describe two topical systems are those of layered graphene and topological insulators. The low-energy Hamiltonian of both a topological insulator surface state (graphene) in the spin (sublattice) basis can be written as \cite{tonyross}

\beq
H(\kvec_\parallel) = \frac{k^2}{2m}\sigma_0 + v_F (\kappa k_{x(y)}\sigma_y + k_{y(x)}\sigma_x) + \Delta\sigma_z,
\label{Hsurf}
\eeq
where $v_F$ is the Fermi velocity, $m$ is the effective mass, $\Delta$ produces a band-gap of $2\Delta$ at $k=0$, and $\kappa = \pm1$ for the $\pm K$ points in graphene, and $\kappa=1$ for topological insulators. In the case of topological insulators, $\Delta = g_s\mu_BB$ in the presence of a magnetic field \cite{zhangrev}.
For graphene, $m\rightarrow \infty$, and $\Delta$ reflects a sublattice anisotropy \cite{gapped}, which vanishes in the case of pristine graphene. 

From \eq{eps}, the eigenvalues are given by $\epsilon_\alpha(k_\parallel) = \frac{k^2}{2m} + \alpha \sqrt{v_F^2k_\parallel^2 + \Delta^2}$, where $\alpha = \pm1$, and the Berry curvature at the Fermi surface is readily obtained as, (reintroducing $\hbar$) \cite{Mx}

\beq
\Omega_{\alpha,\kappa}(k_F) = \frac{\alpha\kappa \Delta \hbar^2v_F^2}{2(\hbar^2v_F^2k_F^2 + \Delta^2)^{\frac{3}{2}}}.
\label{OmDC}
\eeq
Strictly, in the limit $\Delta\rightarrow 0$, the Berry curvature becomes a $\delta-$function at $k=0$ \cite{Mx}. In the limit of vanishing gap, the experimental probe introduced here, namely the frequency and angle dependent interlayer magnetoresistance oscillations, are useless. This is because the frequency dependent AMRO measure the Berry curvature \emph{on the Fermi surface}, which will be zero in such a case except at precisely half filling. For gapped graphene, there are two Fermi circles, one each near the $\pm K$ points. These have opposite Berry curvatures, since $\kappa = \pm1$ for the two valleys.  

As an example of a possible experimental situation, we consider a three dimensional topological insulator thin film (or equivalently a layered two dimensional topological insulator), with interlayer spacing $c=50$ nm, and a Fermi velocity $v_F = 5\times10^5$ ms$^{-1}$, appropriate to Bi$_2$Se$_3$ \cite{hasan}. We consider a Fermi momentum of $k_F = 5\times 10^7$ m$^{-1}$, a g-factor $g_s = 5$, and a relaxation rate corresponding to a mean free path of $100$ nm. In \fig{freqfigfull} we have shown the interlayer magnetoresistance oscillations as a function of tilt angle $\theta$ at $B = 20$ T, for systems both with and without a Berry curvature, and with and without an external field, of frequency $\omega = 6.25\times10^{13}$ s$^{-1}$. For systems without a Berry curvature, the extrema occur at a constant angle, independent of magnetic field. At zero frequency, as shown earlier, the angle is altered slightly due to the Berry curvature. Unfortunately, this effect is tiny. For finite frequency measurements however, we see that the situation is  altered. The location of the minima with and without a Berry curvature are very different, and so we see that the Berry curvature at the Fermi surface leaves a distinct fingerprint in the angle dependent magnetoresistance oscillations. The traces shown in \fig{resfig} are for the system outlined here, in the upper pane the Berry curvature contribution has been subtracted off, and in the lower pane the full result is shown.

In conclusion, we have shown that the presence of a non-zero Berry curvature alters the frequency and angle dependent magnetoresistance oscillations. These results suggest a robust method to measure the Berry curvature at the Fermi surface of topological insulators and superconductors, graphene-like systems, Chern insulators, and topological superconductors. 

\acknowledgments
We thank A. Taskin and Y. Ando for helpful discussions and sharing their experimental data, as well as Omri Bahat-Treidel for a critical reading of the manuscript. ARW is financially supported by a University of Queensland Postdoctoral Fellowship.

\end{document}